\documentclass[preprint]{aastex}
\makeatletter

\def\vereq#1#2{\lower3pt\vbox{\baselineskip1.5pt \lineskip1.5pt
\ialign{$\m@th#1\hfill##\hfil$\crcr#2\crcr\sim\crcr}}}
\makeatother

\begin{document}

\title{The Relationship Between the Circular Polarization and the Magnetic
Field for Astrophysical Masers with Weak Zeeman Splitting}

\author{ W. D. Watson and H. W. Wyld}
\affil{Physics Department, University of Illinois at Urbana-Champaign, 1110
West Green Street, Urbana, IL 61801-3080}

\begin{abstract}
The relationship between the magnetic field and the circular polarization
of astrophysical maser radiation due to the Zeeman effect under 
idealized conditions 
is investigated 
when the Zeeman splitting is much smaller than the spectral linebreadth and
when radiative saturation is significant. The description of the circular
polarization, as well as inferences about the magnetic field from the
observations, are clearest when the rate for stimulated emission is much
less than the Zeeman splitting. The calculations here are performed in this
regime, which is relevant for some (if not most) observations of
astrophysical masers. We
demonstrate that Stokes-$V$ is proportional to the Zeeman
splitting and the fractional linear polarization is independent of the
Zeeman splitting when the ratio of the Zeeman splitting to the spectral
line breadth is small---less than about 0.1. In contrast to its behavior for
ordinary spectral lines, the circular polarization for masers that are at least 
partially saturated does not decrease with 
increasing angle between the magnetic field and the line-of-sight until 
they are nearly perpendicular.
\end{abstract}


\section{INTRODUCTION}

The circular polarization of maser radiation where the Zeeman splitting is
much smaller than the spectral linebreadth is utilized in efforts to obtain
information about the magnetic field in various astronomical contexts
(e.g., for recent work, Kemball \& Diamond 1997; Sarma, Troland, \& Romney
2001; 
Vlemmings, Diamond, \& van Langevelde 2001; Yusef-Zadeh et al. 1999).
For a two-level, non-masing (``thermal") spectral line, a simple relationship
exists
 
\begin{equation}
V/(\partial I/\partial v)=pB \cos\theta
\end{equation}

\noindent that involves Stokes-$V$, the derivative of the intensity $I$
with respect to Doppler velocity $v$, the magnetic field $B$, and the angle
$\theta$  between the magnetic field and the line-of-sight ($p$ is a
constant of the molecular physics for the specific transition). When maser
saturation is unimportant, the relationship can readily be seen as
applicable for masers as well (e.g., Fiebig \& Gusten 1989). Though maser 
polarization in the presence of saturation was the focus of the classic
investigation of Goldreich, Keeley \& Kwan (1973; hereafter GKK), they 
provide no guidance about the circular polarization for weak Zeeman splitting.
GKK only consider line center, where the circular polarization is zero in the 
weak splitting regime. For a summary of the theory of maser polarization,
see Watson (2001).

The GKK idealization is a uniformly pumped, linear maser in steady-state.
The seed radiation is external, weak continuum radiation and the
masing involves an angular momentum $J=1-0$ transition in the presence of a
constant magnetic field. Even though the actual circumstances may be more
complicated, the idealization and solutions of GKK have served as the basis
for discussions of maser polarization in general. In this Letter, we
partially remedy the omission of GKK for weak splitting by calculating
Stokes-$V$ for this basic idealization in the limit that the rate $R$ for
stimulated emission is much less than the Zeeman splitting $g\Omega$
(where $\Omega=eB/m_ec$ and $g$ is the Lande $g$-factor). This
limit certainly is applicable when $g\Omega$ is a noticeable fraction
(greater than a few percent) of the linewidth as occurs for the 1720 MHz OH
masers. It probably also is relevant for the 22 GHz water masers for which
$g\Omega \simeq 10^4B(G)s^{-1}$ for the $F=7-6$ transition and provides at
least a basis for discussion for the SiO masers where $g\Omega\simeq
10^3B(G)s^{-1}$. As a benchmark, the measure of the degree of saturation 
is the ratio of $R$ to the loss rate $\Gamma$ which ordinarily is taken to be 
about 1~s$^{-1}$ and 5~s$^{-1}$ for the H$_2$O and SiO masers, respectively, and 
somewhat less for the 1720 MHz masers. We limit our attention to 
$g\Omega\gg\ R$ because here the
modifications to the standard Zeeman effect are moderate and the magnetic
field can still be inferred from Stokes-$V$ with some confidence. Both the
conceptualization and calculation of Stokes-$V$ are considerably simpler in
this regime because only ``ordinary rate equations" and ``ordinary
molecular populations" of the magnetic substates are involved. In contrast,
when $g\Omega\gg R$  is not satisfied, Stokes-$V$ can be much larger
(sometimes referred to as  ``non-Zeeman" effects) than expected from the
standard Zeeman relationship (Nedoluha \& Watson 1994). It is unclear
whether useful information about the magnetic field can be inferred from
the observations in this regime. The more involved methods of the quantum
mechanical density matrix must be utilized in this regime to incorporate
the correlations between the magnetic substates. Large, non-Zeeman circular
polarization can also be generated by effects that would not be present in
the GKK idealization for masing, such as changes in the direction of the
magnetic field within the masing region (Wiebe \& Watson 1998). 

Calculations of the type being presented here have previously
been performed (Nedoluha \& Watson 1992), but only for a quite limited
range of saturation and angles $\theta$, and only for the molecular
parameters that are specific to the 22 GHz masing transition. That is, for
the high angular momenta ($F=7$, 6, 5, \& 4) and when the values of
the magnetic moments are different for the upper and lower molecular states. 
These calculations did
demonstrate that maser saturation alters equation (1) and that its effect
is to increase Stokes-$V$ much more at large angles than at small angles
relative to that expected from equation (1). The latter is especially
significant because lines-of-sight to masers are often likely to be nearly
perpendicular to the magnetic fields (e.g., toward circumstellar masing rings,
toward the centers of accretion disks viewed edge-on).  The intent of this
Letter is to provide a comprehensive description of the relationship
between the circular polarization and the magnetic field within the
limitations outlined above, and to do this for the idealized masing
conditions treated by GKK which have served as a basis for discussing maser
polarization.

\section{CALCULATIONS}

The rate equations for the normalized (dimensionless) population 
differences $n_+, n_-,$
and $n_0$ between the magnetic substates m = +1, -1, and 0 of the upper
(angular momentum $J=1$) energy level and the m = 0 substate of the lower
($J=0$) level, respectively, are functions of the molecular velocity $v$ and
can be expressed by (e.g., Wallin \& Watson 1995; also GKK)

\begin{equation}
1\ =\ (1+2R_+)n_+\ +\ R_0n_0\ +\ R_-n_-
\end{equation}

\noindent and the two related equations that are obtained by exchanging the
$\pm$
and 0 indexes in equation (2). ``Phenomenological" pumping and loss rates
``$\Lambda$" and ``$\Gamma$", which are standard in maser theory, have been
utilized and incorporated into the normalizations of the populations to
obtain equation (2).  For a unidirectional linear maser, the normalized 
rates for stimulated emission in equation (2) are

\begin{equation}
R_\pm\ =\ I_\pm\ (1+\cos^2\ \theta)\ +\ Q_\pm \sin^2\ \theta\ \pm\ 2V_\pm
\cos\theta
\end{equation}

\noindent and

\begin{equation}
R_0\ =\ 2(I_0 - Q_0)\sin^2\theta
\end{equation}

\noindent where the subscripts $\pm$ and 0 indicate that the intensities
are to be evaluated at the frequencies (angular) $\omega_\pm$ and
$\omega_0$ for which

\begin{equation}
(1-v/c)\omega_\pm\ =\ \omega_R\pm g\Omega/2
\end{equation}

\noindent and

\begin{equation}
(1-v/c)\omega_0=\omega_R
\end{equation}

\noindent Here, $\omega_R$  is the resonance frequency of the transition
and $\pm g\Omega/2$  are the Zeeman shifts of the $m = \pm 1$
magnetic substates. The dimensionless Stokes intensities ($I, Q, V$) are
actual intensities divided by the characteristic saturation intensity $I_S\
=\ (8\hbar \omega^3\Gamma/3\pi c^2A_E)$  where $A_E$ is the Einstein
A-value for the transition. Note that whereas $R_\pm$ and $R_0$ are dimensionless, 
$R$ itself is not normalized and retains its usual dimensions (time$^{-1}$)
in our discussions.
The radiative transfer equations 

\begin{equation}
dI/ds\ = AI\ +\ BQ\ +\ CV
\end{equation}

\begin{equation}
dQ/ds\ =\ AQ + BI
\end{equation}

\begin{equation}
dV/ds = AV+CI
\end{equation}

\noindent are at a specific frequency $\omega$ and can be solved
numerically as a function of the normalized distance $s$ with the 
expressions

\begin{equation}
A=(1+\cos^2\theta)(f_+n_++f_-n_-)+2f_0n_0\sin^2\theta
\end{equation}

\begin{equation}
B=\sin^2\theta(f_+n_+ + f_-n_  -\ -2f_0n_0)
\end{equation}

\begin{equation}
C=2\cos\theta (f_+n_+\ -\ f_-n_-)
\end{equation}

\noindent where the $f$'s are Maxwellian distributions for the component of
the velocity that is along the path of the radiation.  In these equations,
$f_\pm, n_\pm , f_0$ and $n_0$  are evaluated at velocities $v_\pm$ and
$v_0$  given by $(1-v_\pm/c)\omega =\omega_R\pm g\Omega /2$ and
$(1-v_0/c)\omega =\omega_R$.

Starting from the initial conditions $I \ll 1$ and $Q=0=V$, the
radiative transfer equations are integrated numerically at a sufficient
number of frequencies $\omega$ to delineate the profile of the spectral
line. We emphasize that the foregoing equations are valid only when
$g\Omega \gg R$. They are the same as would describe the polarization 
of ordinary thermal radiation---the only differences being the sign of the
net pumping
rate which is reflected in the sign of the ``1" on the left-hand-side of
equation (2) and the ignoring of spontaneous emission.

The main results are presented in the Figure 1 in terms of the ratios
$V/(pB\partial I/\partial v)$ and $Q/I$ computed for a value of $g\Omega$ 
that is much less 
than the spectral linebreadth and at a frequency that is essentially (though not exactly)
at line center. These ratios are useful because, in the weak splittng 
regime and consistent with equation (1), 
they tend to be independent of $g\Omega$ 
and constant with frequency across the spectral line. 
At frequencies where the bulk of the intensity occurs, we have confirmed that the ratios in
Figure 1 are
independent of frequency to a good approximation in the weak splitting regime as delineated in the following 
paragraph. Exceptions to this generalization do occur at the angles for which
the deviations from equation (1) are greatest and also the intensities
are in the neighborhood of $I$=1. Where there are
variations in $V/(pB\partial I/\partial v)$ and $Q/I$ with frequency, 
these variations are modest and are smallest
near line center. Note that the properties of the radiation
at different frequencies within the spectral line are not completely independent
because of the coupling that is indicated in 
equations (5) and (6), and in the 
analogous expressions for $v_\pm $ and $v_0$. 
In detail, the 
rate for stimulated emission depends on the
polarization of the radiation, on the angle $\theta$, and is not exactly the same
for all of the magnetic substates. It nevertheless is useful to utilize $I$
(which is normalized by the characteristic saturation intensity) as the
measure of the degree of saturation. That is, $I\simeq R/\Gamma$.

     The goal here is limited to providing information on the polarization
characteristics for the limit of weak Zeeman splitting---that is, for small
enough $g\Omega $ that the ratios in Figure 1 are independent of $g\Omega$ 
to a good approximation for the bulk of the radiation within the spectral line. 
The precise accuracy of this approximation for a specific $g\Omega$
depends upon angle $\theta$, the degree of saturation, the Doppler velocity
within the spectral line, and is not exactly the same for the linear and
the circular polarization. Nevertheless, some useful generalizations can be
made. We compute polarizations for $g\Omega $ as large as one-fifth of the FWHM
thermal Doppler breadth $\Delta\omega_t$ ($\Delta\omega_t = 2.4\omega_R [kT/Mc^2]^{1/2}$ ).
As long as the degree of saturation is greater than
one, the ratios presented in the Figure 1 are independent of $g\Omega$ 
to an accuracy of a few percent. 
Likewise, as long as $g\Omega /\Delta\omega_t \leq 0.05$,
these ratios also are independent of $g\Omega$ to a similar accuracy regardless of the degree of
saturation. 
For lower saturations---$I$ down to
0.01--- and $0.2 \geq g\Omega/\Delta\omega_t \geq 0.05$ , the ratio
$V/(pB\partial I/\partial
v)$ is still independent of $g\Omega $ to within about twenty percent
accuracy. The main deviations occur at angles less than about forty-five
degrees where the effects of saturation are the least. 

The main effect of
saturation on the circular polarization is readily evident in the Figure 1.
Instead of decreasing as $\cos\theta$ when the angle $\theta$ increases from
zero, Stokes-$V$ tends to increase until a relatively large angle is
reached whose value is determined by the exact degree of saturation. At
$\theta$=0, Stokes-$V$ becomes equal to its value for unsaturated
masing. The nearly discontinuous change reflects the change in the
molecular populations
 that is associated with the similarly rapid variation of the linear
polarization as $\theta$
approaches zero in the GKK theory.
 The variation of the linear
polarization with saturation can be sensitive to the angular momentum of
the molecular states (e.g., Deguchi \& Watson 1990). We have thus performed
similar calculations for a $J=2-1$ masing transition, as well, with the idealization
that $g$ is the same for the upper and lower energy levels (hence, $p=ge/2m_e\omega_R$ 
is the same for both transitions). These also are presented
in Figure 1, along with additional results from a further calculation in which the
$J=1-0$ 
masing is treated as bidirectional.  
The optical depths are the same for radiation propagating in opposite directions in a
linear maser. If the external seed radiation is similar at both ends of the
maser (or if the seed radiation is due to spontaneous emission), the
radiation propagating in the two directions will also be similar and the
maser will be ``bidirectional". It is known that the variation of the
linear polarization with
saturation can be different for uni- and bidirectional masers (Western
\& Watson 1984). However, the comparisons in Figure 1 demonstrate that the
differences for the circular polarization are small for both of these
modifications which 
we have considered beyond the basic unidirectional $J=1-0$ maser .

 The fractional linear polarization in Figure 1 is identical with that
obtained in
previous calculations.  We
present it here in the same format as the circular polarization for
convenience in relating the two polarizations. At high saturation, the
fractional linear polarizations in our calculations (see also Western \& Watson
1984) do reach those of GKK---which are obtained by GKK only for the limit
of high
saturation. The fractional linear
polarization is seen in Figure 1 to increase more rapidly as a function
of saturation for $J=1-0$ than for $J=2-1$ masers, and more rapidly for
bidirectional than for unidirectional masers. As found previously, however, 
to reach the very highest
$Q/I$ given by GKK, the degree of saturation must be implausibly high 
for $J=2-1$ masers.
Anisotropic (or ``$m$-dependent") pumping by starlight is thus the most
likely cause for the highest fractional linear polarizations of the SiO and
perhaps other masers (Western \& Watson 1983). 

 Unfortunately, the degree of radiative saturation of astrophysical masers
is a longstanding uncertainty. In addition to the surface brightness of the
maser, the saturation depends on the angle into which the radiation is
beamed---a quantity for which direct estimates ordinarily are unreliable (e.g.,
Watson \& Wyld 2000). Note that the ``intensities'' of the linear maser formulation
(by us and by GKK) actually are ``mean intensities''$\times 4\pi$. Hence,
intensities from the observations must be multiplied by the solid angle of 
the beam (as well as divided by $I_S$) in order to relate them to the $I$ in the Figures.
A comparison of the observed linear polarization with
that in Figure 1 can be helpful to restrict the degree of saturation if
there are no contaminating effects such as anisotropic pumping, multiple
components, or Faraday rotation. The narrowing of the spectral line during
unsaturated maser
amplification, followed by rebroadening of the line when the maser becomes
saturated, can also be used to obtain an indication of saturation (e.g.,
Nedoluha \& Watson 1991) if there are no 
velocity gradients or multiple components within the maser. A consideration of the linebreadths of
certain prominent 22 GHz water masers indicates that the influence of
saturation on
Stokes-$V$  probably is modest for these masers (Nedoluha \& Watson 1992).
The linebreadth in our calculations is given in Figure 2 as a function of
$I$. It does depend somewhat on the intensity of the incident continuum
seed radiation. 
The computations in Figure 1 are performed for an incident intensity 
$I=10^{-9}$  which we believe to be representative. We have also performed 
computations analogous to those in Figure 1 when the incident intensity $I$ is 
$10^{-5}$ as might occur when the masing gas is amplifying a strong continuum
source. The resulting $Q/I$ are 
essentially identical to those in Figure 1. The Stokes-$V$  also are quite similar to those in 
Figure 1, except near the peaks of the curves for the the largest $I$ ($\ga
10^2$)
where they are smaller by 15--20\%.

 In summary, when the observed masers are believed to be at least
somewhat saturated, but there is no good information about the angle
$\theta$ nor about the exact degree of saturation, simply removing the
$\cos\theta$ in equation (1) would seem to provide the best way at present to infer
magnetic field strengths from the observed Stokes-$V$ in the weak splitting regime 
when ``non-Zeeman effects" can be ignored. Saturation with $I\ga 10^2$ (and probably
even $I\ga 10$) seems unlikely for astrophysical masers. In contrast to 
the linear polarization, the circular polarization is relatively insensitive
to the angular momentum of the molecular states. We emphasize that our results
are applicable in detail only for the idealized masing conditions on which the
calculations are based (see Introduction). In addition to non-Zeeman effects,
velocity gradients, anisotropic
pumping, and multiple hyperfine components may be present, but are not considered here.
For example, the 22 GHz masing transition of water 
probably consists of multiple hyperfine components
so that equation (1) is unlikely to be directly applicable for it (but see
Nedoluha \& Watson 1992).

This research has been supported in part by NSF Grant AST99-88104

\noindent REFERENCES
\begin{description}
\item{Deguchi, S., \& Watson, W.D. 1990, ApJ, 354, 649}
\item{Fiebig, D., \& Gusten, R. 1989, A\&A, 214, 333}
\item{Goldreich, P., Keeley, D.S., \& Kwan, J.Y. 1973, ApJ, 179,111 [GKK]}
\item{Kemball, A. J., \&  Diamond, P. J.. 1997, ApJ, 481, L111}
\item{Nedoluha, G. E., \& Watson, W. D. 1991, ApJ, 367, L63}
\item{Nedoluha, G. E., \& Watson, W. D. 1992, ApJ, 384, 185}
\item{Nedoluha, G. E., \& Watson, W. D. 1994, ApJ, 423, 394}
\item{Sarma, A. P., Troland, T. H., \& Romney, J. D. 2001, ApJ, 554, L217}
\item{Vlemmings, W., Diamond, P. J.,\& van Langevelde, H. J. 2001, in 
proceedings of IAU Symposium 206: Cosmic Masers, in press}
\item{Wallin, B.K., \& Watson, W.D. 1995, ApJ, 445, 465} 
\item{Watson, W. D. 2001, in proceedings of IAU Symposium 206: 
Cosmic Masers, in press}
\item{Watson, W.D., \& Wyld, H. W. 2000, ApJ, 530, 207} 
\item{Western, L.R., \& Watson, W. D. 1983, ApJ, 275, 195}
\item{Western, L.R., \& Watson, W. D. 1984, ApJ, 285, 158}
\item{Wiebe, D.S., \& Watson, W. D. 1998, ApJ, 503, L71}
\item{Yusef-Zadeh, F., Roberts, D. A., Goss, W. M., Frail, D. A., \& Green,
A. J.
1999, ApJ, 512,230}
\end{description}
\clearpage

\figcaption{Circular and linear polarization of maser radiation as a 
function of the cosine of the angle $\theta$ between the direction 
of the magnetic field and the line of sight. Polarizations are 
presented for $J=1-0$ unidirectional and bidirectional masers, 
and for $J=2-1$ unidirectional masers, in separate panels as 
indicated by the label in 
each panel.The curves are labeled 
by the $\log_{10}$ of the intensity $I$ which is essentially the degree 
of saturation. Curves (some of which overlap) are plotted for
$I=10^{-2}$, $10^{-1}$, 1, 3, 10, $10^2$, $10^3$, and $10^4$.\\
(upper three panels) The circular polarization as measured by the 
magnitude of
$V/(pB\partial I/\partial v)$ which is equal to $\cos\theta$ in the unsaturated
limit.\\
(lower three panels) The fractional linear polarization. Stokes-$Q$ 
is positive when the linear polarization is perpendicular to the direction
of the
magnetic field projected onto the plane of the sky.}

\figcaption{The ratio of the spectral linebreadth (FWHM) of the maser radiation
to the Doppler breadth $\Delta\omega_t$ (FWHM) of the thermal velocities
of the masing molecules, as a function of the $\log_{10}$ of the intensity $I$. Linebreadths 
are shown for $J=1-0$ unidirectional
(solid lines), $J=1-0$ bidirectional (dotted lines), $J=2-1$ unidirectional
(dashed lines)
masers with intensities for the incident continuum radiation of 
$I=10^{-5}$, as well as for $I=10^{-9}$ which we consider to be 
most representative.}
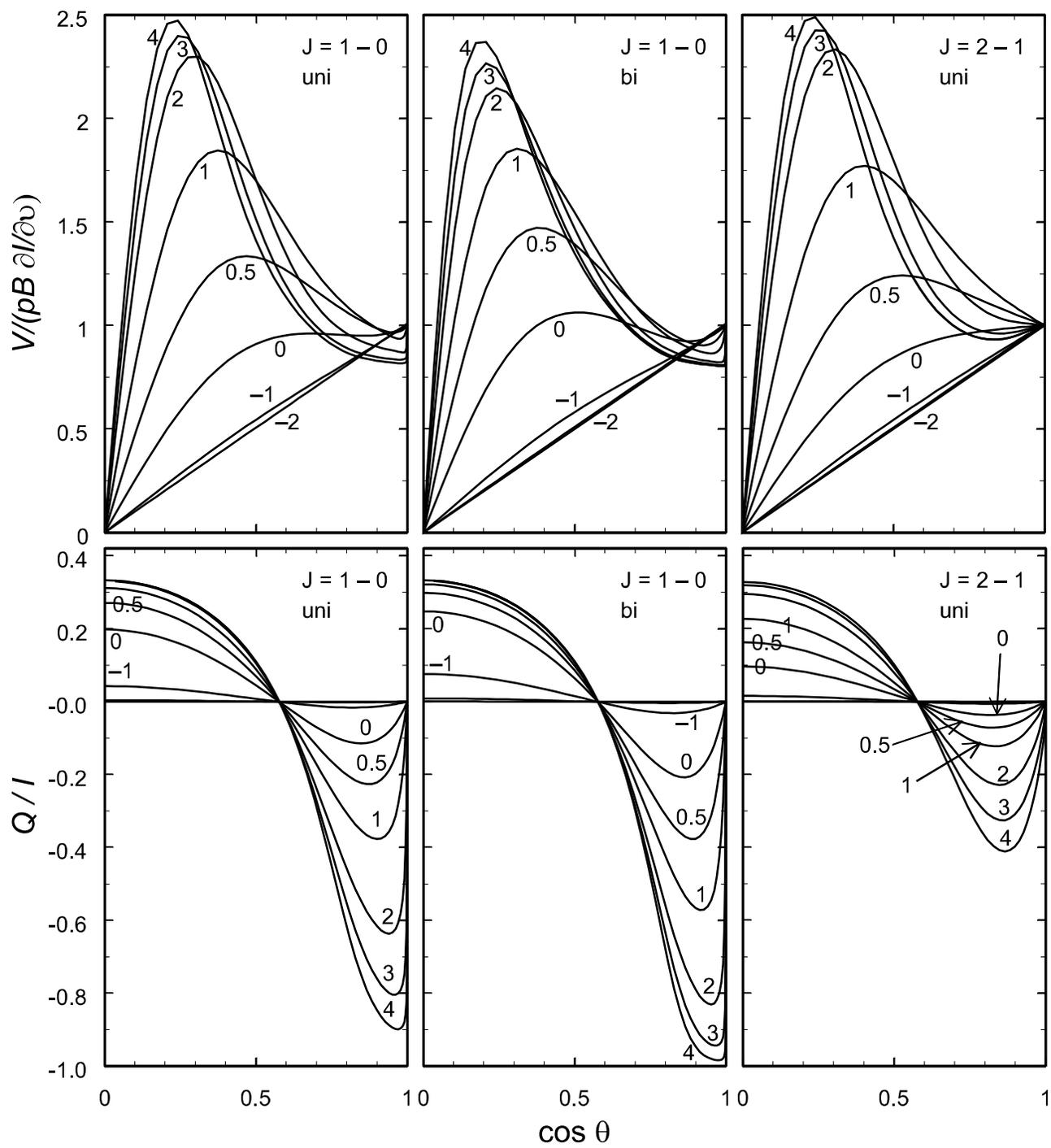
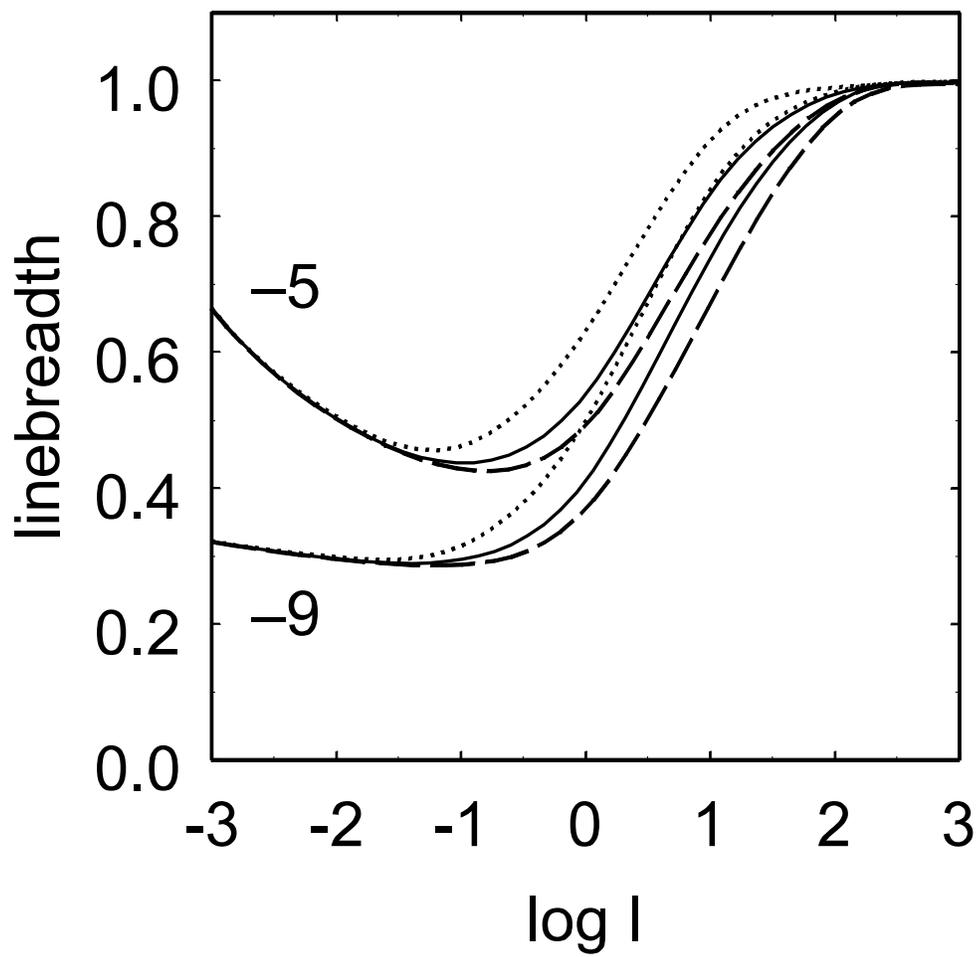
\end{document}